\documentclass[12pt]{article}
\usepackage{amssymb,latexsym,graphicx,a4,epsfig,psfrag,here}

\newcommand{\bda}{\begin{\displaymath}\begin{array}{rl}}
\newcommand{\eda}{\end{array}\end{displaymath}}
\newcommand{\be}{\begin{equation}}
\newcommand{\ee}{\end{equation}}
\newcommand{\bdm}{\begin{displaymath}}
\newcommand{\edm}{\end{displaymath}}
\newcommand{\bea}{\begin{eqnarray}}
\newcommand{\eea}{\end{eqnarray}}
\newcommand{\no}{\nonumber \\}

\def\Mpin{M_{\pi^0}^2}
\def\mpin{M_{\pi^0}}
\def\Mpic{M_{\pi^{\pm}}^2}
\def\mpic{M_{\pi^{\pm}}}

\def\MKc{M_{K^{\pm}}^2}
\def\mKc{M_{K^{\pm}}}

\def\Fpi{F_{\pi}^2}
\def\fpi{F_{\pi}}

\def\Jpm{{\bar J}_{+-}}
\def\Joo{{\bar J}_{00}}
\def\kbar{{\bar k}}

\begin{document}

\thispagestyle{empty}
\begin{flushright}
CPT-2001/P.4271 \\
\today
\end{flushright}

\vspace{2cm}

\begin{center}{\Large {\bf {Electromagnetic Corrections to Charged Pion Scattering at Low Energies}}}

\vspace{0.5cm}

M.~Knecht\footnote{knecht@cpt.univ-mrs.fr} and A.~Nehme\footnote{nehme@cpt.univ-mrs.fr}

\vspace{0.3cm}
\begin{center}
\textit{Centre de Physique Th\'eorique} \\
\textit{CNRS-Luminy, case 907} \\
\textit{F-13288 Marseille Cedex 09, France}.
\end{center}

\vspace{2cm}


\begin{abstract}
The electromagnetic corrections to the low energy scattering amplitude 
involving charged pions only are investigated 
at leading and next-to-leading orders in the two-flavour chiral expansion. 
As an application, the corresponding variation in the strong $2S-2P$ 
level shift is evaluated. The relative variation is of the order of $5\%$.
\end{abstract}



\end{center}
\newpage
\renewcommand{\baselinestretch}{0.6}

\renewcommand{\baselinestretch}{1}


\paragraph{}

{\bf {1.}}~The study of hadronic atoms has become a very active field
(see \cite{Gasser:2001iz} for a recent account). 
Many experiments are devoted to measure the characteristics of such 
atoms with high 
precision~\cite{Adeva:1994xz,Sigg:qe,Breunlich:af,Iwasaki:1997wf,Adeva:2000vb}.
 These experimental results carry a well 
founded theoretical interest, since they provide a direct access to hadronic 
scattering lengths, leading, in this way, to valuable 
informations concerning the fundamental properties of QCD at low energy. 
For instance, the 
presently running DIRAC experiment aims at measuring the pionium lifetime 
$\tau$ with $10\%$ accuracy~\cite{Adeva:1994xz}. This would allow 
one to determine the 
difference $a_0^0-a_0^2$ with $5\%$ precision by means of the 
Deser-type~\cite{Deser:1954vq} relation~\cite{Bilenkii:zd} 
\be \label{eq:lifetime}
\tau^{-1}\propto\,\left (a_0^0-a_0^2\right )^2\,, 
\ee
where $a_l^I$ denotes the $l$-wave $\pi\pi$ scattering length in the channel 
with total isospin $I$. On the other hand, chiral perturbation 
theory (ChPT) predictions for the scattering lengths have reached a precision 
amounting to 
$2\%$~\cite{Colangelo:2000jc}. Before confronting the experimental 
determination to the ChPT prediction, 
it is necessary to get all sources of corrections to the relation 
(\ref{eq:lifetime}), valid in the absence of isospin 
breaking, under control. In this connection, bound state calculations were 
performed using different approaches, like potential scattering 
theory~\cite{Moor:ye,Gashi:1997ck}, 
$3D$-constraint field theory~\cite{Jallouli:1997ux}, Bethe-Salpeter 
equation~\cite{Lyubovitskij:1996mb} and non-relativistic effective 
lagrangians~\cite{Kong:1998xp, Gasser:2001un}. For a review on 
the subject and a 
comparison between the various methods we refer the reader 
to~\cite{Gasser:1999vf}. Within the framework of non-relativistic effective 
lagrangians, the 
correct expression of relation (\ref{eq:lifetime}) which include all 
isospin breaking effects at leading order (LO) and next-to-leading order 
(NLO) was given as~\cite{Gall:1999bn} 
\be \label{eq:corrected lifetime}
\tau^{-1}\,=\,\frac{1}{9}\,\alpha^3\left (4\Mpic -4\Mpin -\Mpic\,\alpha^2\right )^{\frac{1}{2}}{\cal A}^2\,(1+K)\,.
\ee
In the preceding equation, $\alpha =e^2/(4\pi )$ stands for the fine-structure constant, ${\cal A}$ and $K$ possess the following 
expansions~\cite{Gall:1999bn} in powers 
of the isospin breaking parameter $\kappa\in \left [\alpha ,\,(m_d-m_u)^2\right ]$
\bea
{\cal A} &=& -\frac{3}{32\pi}\,\textrm{Re}~A_{\textrm{\scriptsize thr.}}^{+-;00}+o(\kappa )\,, \\ 
K        &=& \frac{1}{9}\left (\frac{\Mpic}{\Mpin}-1\right )\left (a_0^0+2a_0^2\right )^2-\frac{2\alpha}{3}\,(\ln\alpha -1)
               \left (2a_0^0+a_0^2\right )+o(\kappa )\,. 
\eea
The quantity of interest, 
\be \label{eq:charged to neutral}
-\frac{3}{32\pi}\,\textrm{Re}~A_{\textrm{\scriptsize thr.}}^{+-;00}\,=\,a_0^0-a_0^2+h_1(m_d-m_u)^2+h_2\alpha\,,
\ee 
represents the real part of the $\pi^+\pi^-\rightarrow\pi^0\pi^0$ scattering amplitude at order $\kappa$, calculated at threshold within 
ChPT to any 
chiral order and from which we subtract the singular pieces behaving like $q^{-1}$ and $\ln q$ whith $q$ being the center-of-mass 
three-momentum of the charged pions. The coefficient $h_2$ was calculated in~\cite{Knecht:1997jw} at next-to-leading 
order in the chiral expansion while $h_1\,=\,{\cal O}(m_u+m_d)$~\cite{Gasser:1999vf}.

The DIRAC proposal~\cite{Adeva:1994xz} also mentioned the possibility to measure the strong $2S-2P$ energy level shift 
$\Delta E_{\textrm{\scriptsize strong}}$ of the pionium. How this measurement could be performed in practice has been discussed 
in Ref.~\cite{Nemenov:vp}. A simultaneous measurement of $\tau$ and $\Delta E_{\textrm{\scriptsize strong}}$ would allow to pin 
down $a_0^0$ and 
$a_0^2$ separately, since~\cite{Efimov:1985fe} 
\be \label{eq:energy levels pionium}
\Delta E_{\textrm{\scriptsize strong}}\propto\,2a_0^0+a_0^2\,.
\ee
Bound state calculation of the isospin breaking corrections to (\ref{eq:energy levels pionium}) were done in~\cite{Gashi:1997ck} using 
potential scattering theory (the main contribution to the \underline{total} level shift comes from vacuum polarization effects, see e.g.~\cite{Eiras:2000rh}).
Non-relativistic effective lagrangian calculations concerning 
$\Delta E_{\textrm{\scriptsize strong}}$ are not available for pionium, but 
exist for the pionic hydrogen case~\cite{Lyubovitskij:2000kk}. 
One may thus reasonably expect an expression similar to 
Eq. (\ref{eq:corrected lifetime})
\bea
\Delta E_{\textrm{\scriptsize strong}} 
  &=& -\frac{1}{8}\,\alpha^3\mpic{\cal A}'(1+ K')\,, \no 
{\cal A}'
  &=& \frac{1}{32\pi}\,\textrm{Re}~A_{\textrm{\scriptsize thr.}}^{+-;+-}+o(\kappa )\,,
\eea 
involving the $\pi^+\pi^-\rightarrow\pi^+\pi^-$ scattering amplitude $A^{+-;+-}$. Electromagnetic corrections to this scattering amplitude will be 
calculated at next-to-leading order in the present work. 



\paragraph{}

\begin{figure} 
\begin{center}
\includegraphics{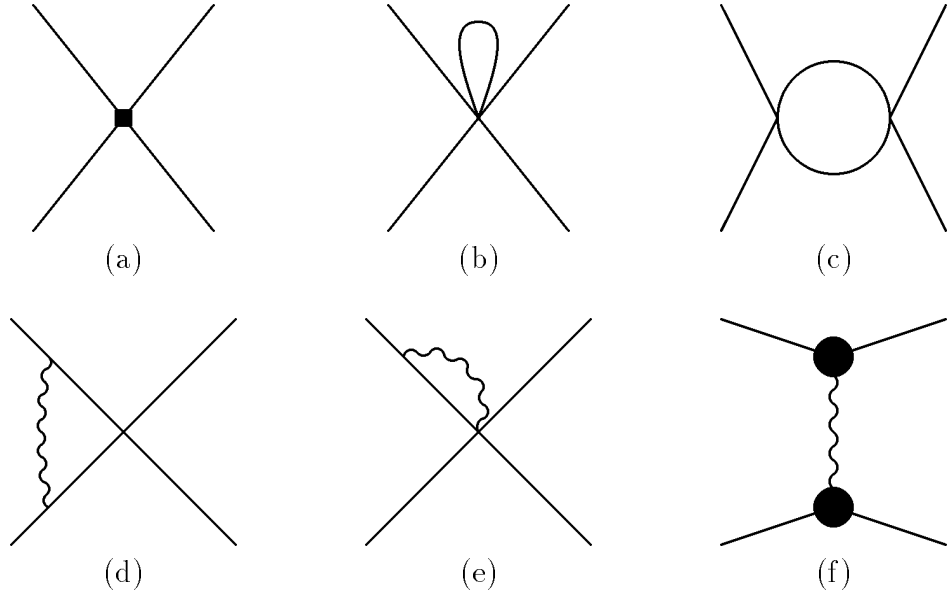}
\caption{\label{fig:2}The various topologies of Feynman diagrams contributing 
to the charged $\pi\pi$ scattering amplitude at order one 
loop, but ignoring ${\cal O}(e^4)$ effects. The full circles 
appearing in the Born-type diagram (f) are made explicit in Fig.~\ref{fig:3}.}
\end{center}
\end{figure}

\begin{figure} 
\begin{center}
\includegraphics{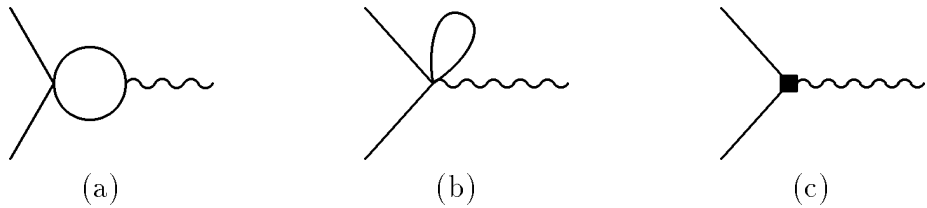}
\caption{\label{fig:3}The electromagnetic vertex function of a charged pion 
to one-loop order. The full square takes into account the 
contribution from the low-energy constants just as the tree contribution 
including the effect of wave function renormalization. Diagrams 
of order ${\cal O}(e^3\,p)$ are discarded.} 
\end{center}
\end{figure}

{\bf {2.}}~The elastic scattering process 
\be \label{eq:process}
\pi^+(p_+)\,+\,\pi^-(p_-)\,\rightarrow\,\pi^+(p_+')\,+\,\pi^-(p_-')\,,
\ee
is studied in terms of the Lorentz invariant Mandelstam variables
$$
s=(p_++p_-)^2\,, \qquad t=(p_+-p_+')^2\,, \qquad u=(p_+-p_-')^2\,,
$$
satisfying the on-shell relation $s+t+u\,=\,4\Mpic$. These variables are 
related to the center-of-mass three-momentum $q$ and 
scattering angle $\theta$ by  
\be
s=4(\Mpic +q^2)\,,\quad  t=-2q^2(1-\cos\theta )\,,\quad  
u=-2q^2(1+\cos\theta )\,. \label{eq:c.m.f.}
\ee
Let $A^{+-;+-}$ and $A^{++;++}$ denote the respective scattering amplitudes 
for the process (\ref{eq:process}) and for the 
crossed channel reaction $\pi^+\pi^+\rightarrow\pi^+\pi^+$. Then, 
$s\leftrightarrow u$ crossing is expressed as 
\be \label{eq:direct crossed}
A^{+-;+-}(s, t, u)\,=\,A^{++;++}(u, t, s)\,. 
\ee
We shall calculate the scattering amplitude (\ref{eq:direct crossed}) at NLO 
including electromagnetic effects. The strong sector chiral 
lagrangian for two-flavour ChPT was constructed in~\cite{Gasser:1983yg}. 
Treating isospin violation of electromagnetic origin requires the 
extension of ChPT in order to include virtual photons. This can be done by 
building operators in which photons figure as explicit dynamical degrees of 
freedom. The electromagnetic sector of the chiral lagrangian for two-flavour 
ChPT has been discussed 
at NLO in~\cite{Knecht:1997jw} and~\cite{Meissner:1997fa}. We shall work in 
the $m_u=m_d$ limit and use the lagrangian representation 
of~\cite{Knecht:1997jw}. At one-loop accuracy, all of the following chiral 
orders 
are present: $p^2, e^2, p^4, e^2p^2, e^4$. We do however not consider the 
${\cal O}(e^4)$ contributions, 
which include two-photon exchange box diagrams, and which are expected to be 
smaller than the other contributions at the same order. Using 
Feynman graph techniques, the amplitude (\ref{eq:direct crossed}) can be 
represented at NLO by the  
diagrams\footnote{If one uses the so-called $\sigma$-model 
parametrization of the pion fields, the diagrams 
(e) of Fig.~\ref{fig:2} and (b) of Fig.~\ref{fig:3} vanish identically.} 
depicted in Fig.~\ref{fig:2}. 

The scattering amplitude (\ref{eq:direct crossed}) 
is conveniently written in the following $s\leftrightarrow t$ 
symmetric decomposition
\bea
A^{+-;+-}(s,t,u) 
  &=& \Bigg\{\frac{s-\Mpin}{F^2}+B^{+-;+-}(s,t,u)+C^{+-;+-}(s,t,u) \no 
  &+& e^2\left (\frac{u-t}{s}\right )
\left [F_V^{\pi}(s)\right ]^2\Bigg\}+\{s\leftrightarrow t\}\,. 
\label{eq:scattering amplitude}
\eea
In the preceding expression, $B^{+-;+-}$ collects the unitarity pieces 
arising from the diagrams of type (c) and (d) in Fig.~\ref{fig:2},  
\bea
B^{+-;+-}(s,t,u)
  &=& \frac{1}{2F^4}\,(s-\Mpin )^2\Joo (s) \no 
  &+& \frac{1}{F^4}\left [\frac{s^2}{4}-\frac{1}{12}\,(u-t)(s-4\Mpic )+2s\Delta_{\pi}+4\Delta_{\pi}^2\right ]\Jpm (s) \no 
  &+& \frac{1}{4F^4}\,(u-2\Mpic -2\Delta_{\pi})(u-2\Mpic -2\Delta_{\pi}-4e^2F^2)\Jpm (u) \no 
  &+& \frac{2e^2}{F^2}\,(u-2\Mpic -2\Delta_{\pi})\left [2(s-2\Mpic )G_{+-\gamma}(s)-(u-2\Mpic )G_{+-\gamma}(u)\right ] \no 
  &-& \frac{e^2}{F^2}\left [s+4\Delta_{\pi}-4(s-2\Mpic )\left (\frac{t-u}{t+u}\right )\right ]\Jpm (s)\,, \label{eq:unitary correction}
\eea
where 
$$
\Delta_{\pi}\,=\,\Mpic -\Mpin\,. 
$$
The expressions for the various loop functions appearing in 
(\ref{eq:unitary correction}) can be found in~\cite{Knecht:1997jw}. The 
function $C^{+-;+-}$ represents the contributions from tadpoles as well as 
from the strong~\cite{Gasser:1983yg} and 
electromagnetic~\cite{Knecht:1997jw} low-energy constants
\bea
C^{+-;+-}(s,t,u)
  &=& \frac{s-\Mpin}{F^2}\,\frac{e^2}{32\pi^2}\left [-18-8\left (1+\ln\frac{m_{\gamma}^2}{\Mpic}\right )+
        \frac{1}{2}\left ({\cal K}^{+-;+-}-{\cal K}^{++;++}\right )\right ] \no 
  &+& \frac{e^2\Mpin}{32\pi^2F^2}\left [10+\frac{1}{2}\left ({\cal K}^{+-;+-}+{\cal K}^{++;++}\right )\right ]
        -\frac{e^2}{2\pi^2F^2}\,(s-2\Mpic )\left (\frac{t-u}{t+u}\right ) \no 
  &+& \frac{1}{48\pi^2F^4}\left [(s-2\Mpic )^2({\bar l}_1+{\bar l}_2)+(u-2\Mpic )^2{\bar l}_2\right ]
        -\frac{\mpin^4}{32\pi^2F^4}\,{\bar l}_3 \no 
  &+& \frac{1}{16\pi^2F^4}\left (-\frac{5}{18}\,u^2-\frac{13}{18}\,s^2+\frac{2}{3}\,u\Mpin +
        \frac{19}{6}\,u\Delta_{\pi}+\frac{5}{18}\mpin^4-\frac{58}{9}\,\Mpin\Delta_{\pi}\right ) \no 
  &-& \frac{1}{96\pi^2F^4}\,\frac{\Delta_{\pi}}{\Mpin}\left (-3s^2+16s\Mpin +2u\Mpin -23\mpin^4\right )\,. 
\label{eq:polynomial correction}
\eea
In this last expression, we have dropped all contributions of order 
${\cal O}(e^4)$ and beyond. In particular, we have expanded all logarithms 
of the pion mass ratio,
$$
\ln\frac{\Mpic}{\Mpin}\,=\,\frac{\Delta_{\pi}}{\Mpin}+\ldots \,.
$$
The terms ${\cal K}^{+-;+-}$ and ${\cal K}^{++;++}$ involve combinations
of the electromagnetic counterterms $k_i$,
\bea
{\cal K}^{+-;+-} 
  &=& \left (3+\frac{4Z}{9}\right )\kbar_1-\frac{40Z}{9}\kbar_2-
9\kbar_3+4Z\kbar_4+4(1+8Z)\kbar_6+2(1-8Z)\kbar_8\,, 
        \label{eqa:combination} \\ 
{\cal K}^{++;++} 
  &=& -\left (3+\frac{4Z}{9}\right )\kbar_1-\frac{248Z}{9}\kbar_2+9\kbar_3-
20Z\kbar_4+4(1+8Z)\kbar_6+2(1-8Z)\kbar_8\,. 
        \label{eqb:combination}
\eea 
The low-energy constants ${\bar l}_i$ and ${\bar k}_i$ are related to the 
respective renormalized running couplings 
$l_i^r(\mu )\equiv l_i^r$ and $k_i^r(\mu )\equiv k_i^r$ at the scale 
$\mu =\mpic$,
$$
l_i^r=\frac{\gamma_i}{32\pi^2}\left ({\bar l}_i+
\ln\frac{\Mpic}{\mu^2}\right )\,, 
\qquad k_i^r=\frac{\sigma_i}{32\pi^2}\left ({\bar k}_i+
\ln\frac{\Mpic}{\mu^2}\right )\,,
$$
where $\gamma_i$ and $\sigma_i$ are the renormalization group beta-functions 
whose values are given in~\cite{Gasser:1983yg} 
and~\cite{Knecht:1997jw}, respectively. The last quantity that remains 
to be defined in (\ref{eq:scattering amplitude}) is $F_V^{\pi}$, which 
denotes the electromagnetic pion form factor
$$
\langle\pi^+(p_+')|{\cal J}^{\mu}(0)|\pi^+(p_+)\rangle\,=\,-ie(p_++p_+')^{\mu}F_V^{\pi}(Q^2)\,, \qquad Q=p_+'-p_+\,,
$$
with ${\cal J}^{\mu}$ being the electromagnetic current.
At the order we are working, it is given by~\cite{Gasser:1983yg, Kubis:1999db}  
\be
F_V^{\pi}(p^2)\,=\,1+\frac{p^2}{6F^2}\left [\left (1-\frac{4\Mpic}{p^2}\right )\Jpm (p^2)+\frac{1}{16\pi^2}
\left ({\bar l}_6-\frac{1}{3}\right )\right ]\,. 
\ee
Finally, let us mention that the scattering amplitude (\ref{eq:scattering amplitude}) is scale independent but infrared 
divergent. This infrared divergence is signaled by the presence of a $\ln\,(m_{\gamma}^2/\Mpic )$ term, where $m_{\gamma}$ is a small 
photon mass acting as an infrared regulator. These infrared divergencies have 
to cancel in observable quantities, like the total cross 
section with soft photon emission. These divergencies do however not show up 
in the discussion of the electromagnetic corrections to the 
strong level shift, to which we now turn. 
 


\paragraph{}

{\bf {3.}}~The scattering lengths $a_0^0$ and $a_0^2$ are well-defined 
quantities in the absence of radiative corrections. Their NLO 
expressions were derived in~\cite{Gasser:1983yg} and, for historical 
reasons~\cite{Weinberg:1966kf}, we shall reproduce them in terms of 
the charged pion mass 
\bea 
a_0^0 
  &=& \frac{7\Mpic}{32\pi\Fpi}\left\{1+\frac{5\Mpic}{84\pi^2\Fpi}\left [{\bar l}_1+2{\bar l}_2-\frac{3}{8}\,{\bar l}_3
        +\frac{21}{10}\,{\bar l}_4+\frac{21}{8}\right ]\right\}\,, \no 
a_0^2 
  &=& -\frac{\Mpic}{16\pi\Fpi}\left\{1-\frac{\Mpic}{12\pi^2\Fpi}\left [{\bar l}_1+2{\bar l}_2-\frac{3}{8}\,{\bar l}_3
        -\frac{3}{2}\,{\bar l}_4+\frac{3}{8}\right ]\right\}\,, \label{eq:scattering lengths}
\eea
with $\fpi$ being the pion decay constant whose NLO expression is given in the isospin limit by~\cite{Gasser:1983yg}
$$
\fpi\,=\,F\left (1+\frac{\Mpin}{16\pi^2F^2}\,{\bar l}_4\right )\,.
$$        
We are interested in electromagnetic corrections to the {\it strong} energy 
level shift. To this end, one has to consider the part of the 
amplitude corresponding to one-photon-irreducible (OPI) diagrams. It is then 
convenient to subtract the contribution from diagram (f) of 
Fig.~\ref{fig:2} as follows
\bea
A^{+\mp ;+\mp}(s,t,u) 
  &=& A_{\textrm{\scriptsize OPI}}^{+\mp ;+\mp}(s,t,u)+
A_{\textrm{\scriptsize Born}}^{+\mp ;+\mp}(s,t,u)\,, \no 
A_{\textrm{\scriptsize Born}}^{+-;+-}(s,t,u)
  &=& e^2\left\{\frac{u-t}{s}\,\left [F_V^{\pi}(s)\right ]^2+\frac{u-s}{t}\,
\left [F_V^{\pi}(t)\right ]^2\right\}\,, \no
A_{\textrm{\scriptsize Born}}^{++;++}(s,t,u)
  &=& e^2\left\{\frac{s-t}{u}\,\left [F_V^{\pi}(u)\right ]^2+\frac{s-u}{t}\,
\left [F_V^{\pi}(t)\right ]^2\right\}\,. 
\eea
Next, we expand the real part 
of $A_{\textrm{\scriptsize OPI}}^{+\mp ;+\mp}$ in powers of $q$,
\bea
\textrm{Re}~A_{\textrm{\scriptsize OPI}}^{+-;+-}(s,t,u)
  &=& \frac{\Mpin}{\Fpi}\,\frac{e^2}{4}\,\frac{\mpin}{q}
        +\textrm{Re}~A_{\textrm{\scriptsize thr.}}^{+-;+-}+{\cal O}(q)\,, 
\label{eqa:coulomb} \\
\textrm{Re}~A_{\textrm{\scriptsize OPI}}^{++;++}(s,t,u)
  &=& \frac{\Mpin}{\Fpi}\,\frac{e^2}{4}\,\frac{\mpin}{q}
        +\textrm{Re}~A_{\textrm{\scriptsize thr.}}^{++;++}+{\cal O}(q)\,. 
\label{eqb:coulomb}
\eea
The term in $q^{-1}$ is due to the 
Coulomb photon exchanged between the scattered particles in diagram (d) of 
Fig.~\ref{fig:2}. It should be absorbed in the static 
characteristics of pionium~\cite{Bilenkii:zd, Nemenov:1984cq, Belkov:1986xn} 
within the treatment of the  bound state properties. Furthermore,
the infrared divergent terms show up only in the ${\cal O}(q)$ terms.
The regular terms, $\textrm{Re}~A_{\textrm{\scriptsize thr.}}^{+-;+-}$ and 
$\textrm{Re}~A_{\textrm{\scriptsize thr.}}^{++;++}$, constitute 
the central object of the current work. Their isospin limits are nothing else 
than $2a_0^0+a_0^2$ and $a_0^2$ respectively,
\bea
\frac{1}{32\pi}\,\textrm{Re}~A_{\textrm{\scriptsize thr.}}^{+-;+-}
  &=& \frac{1}{6}\,(2a_0^0+a_0^2)+\Delta a_0(+-;+-)\,, 
\label{eqa:electromagnetic corrections} \\
\frac{1}{32\pi}\,\textrm{Re}~A_{\textrm{\scriptsize thr.}}^{++;++} 
  &=& a_0^2+\Delta a_0(++;++)\,. \label{eqb:electromagnetic corrections}
\eea
The NLO expressions for the electromagnetic corrections are found to be
\bea
32\pi\Delta a_0(+-;+-)
  &=& \frac{2\Delta_{\pi}}{\Fpi}+\frac{\Mpin\Delta_{\pi}}{8\pi^2\fpi^4}
\,(2+{\bar l}_3)-\frac{e^2\Mpin}{16\pi^2\Fpi}\,
        \left (24-{\cal K}^{+-;+-}\right )\,, \label{eqa:expression for 
the electromagnetic corrections} \\
32\pi\Delta a_0(++;++)
  &=& \frac{2\Delta_{\pi}}{\Fpi}+\frac{\Mpin\Delta_{\pi}}{16\pi^2\fpi^4}\,(3+2{\bar l}_3+8{\bar l}_4)-\frac{e^2\Mpin}{16\pi^2\Fpi}
        \left (20-{\cal K}^{++;++}\right )\,, \label{eqb:expression for the 
electromagnetic corrections}
\eea
with the combinations of the electromagnetic low-energy constants 
defined in (\ref{eqa:combination}) and (\ref{eqb:combination}).



\paragraph{}

{\bf {4.}}~For the numerical estimates of Eqs.~(\ref{eqa:expression for the 
electromagnetic corrections}) 
and~(\ref{eqb:expression for the electromagnetic corrections}), 
we use the following values~\cite{Groom:in}. The charged and neutral pions masses are  
$\mpic =139.570\,\textrm{MeV}$ and $\mpin =134.976\,\textrm{MeV}$, 
respectively. The pion decay constant is taken as 
$\fpi =92.4\,\textrm{MeV}$. The values for the 
strong sector low-energy constants are: 
${\bar l}_3=2.9\pm 2.4$~\cite{Gasser:1983yg}, 
${\bar l}_4=4.4\pm 0.3$~\cite{Bijnens:1998fm}. As for the two-flavour 
low-energy constants ${\bar k}_i$, no direct numerical estimates exist up 
to now. On the other hand, some of the corresponding 
three-flavour counterterms $K_i^r(\mu )\equiv K_i^r$ 
were evaluated in~\cite{Baur:1996ya, Moussallam:1997xx, Bijnens:1996kk}. 
In order to give a numerical estimate of the 
combinations (\ref{eqa:combination}) and (\ref{eqb:combination}), it is then 
necessary to relate the $k_i^r$'s to the 
$K_i^r$'s (note that the combinations of 
$k_i^r$'s involved in the amplitude $\pi^+\pi^-\to\pi^0\pi^0$ were expressed 
in terms of the $K_i^r$'s 
in~\cite{Jallouli:1997ux} and~\cite{Gasser:2001un, Gasser:1999vf}). In the 
present case, this requires to compute the corresponding 
three-flavour representations of 
the quantities $\textrm{Re}~A_{\textrm{\scriptsize thr.}}^{+-;+-}$ and 
$\textrm{Re}~A_{\textrm{\scriptsize thr.}}^{++;++}$~\cite{Nehme:2002}, 
and then to match them with 
Eqs. (\ref{eqa:electromagnetic corrections}) and 
(\ref{eqb:electromagnetic corrections}), respectively, in the limit of a very 
massive strange quark. 
With the help of the strong sector matching relations~\cite{Gasser:1984gg} 
between 
the $l_i^r$'s and the $L_i^r$'s, we obtain
\bea
{\cal K}^{+-;+-} 
  &=& \frac{64\pi^2}{9}\left [-12(K_1^r+K_2^r)+9(6K_3^r+K_4^r)-10(K_5^r+K_6^r)+72(K_8^r+K_{10}^r+K_{11}^r)\right ] \no
  &-& 6Z_0\left (1+\ln\frac{B_0m_s}{\mu^2}\right )-16Z_0\left ({\bar l}_4+2\ln\frac{\Mpic}{\mu^2}\right )\,, \label{eqa:matching} \\                
{\cal K}^{++;++}
  &=& \frac{64\pi^2}{9}\left [12(K_1^r-5K_2^r)-9(6K_3^r+5K_4^r)+2(5K_5^r-31K_6^r)+72(K_8^r+K_{10}^r+K_{11}^r)\right ] \no
  &-& 2Z_0\left (1+\ln\frac{B_0m_s}{\mu^2}\right )-16Z_0\left ({\bar l}_4-\ln\frac{\Mpic}{\mu^2}\right )-12\ln\frac{\Mpic}{\mu^2}\,, 
        \label{eqb:matching} 
\eea
where $Z_0$ is the three-flavour analogue of $Z$, 
$$
Z\,=\,\frac{\Delta_{\pi}}{2e^2F^2}\biggm |_{m_u=m_d=0}\,, \qquad  Z_0\,=\,\frac{\Delta_{\pi}}{2e^2F_0^2}\biggm |_{m_u=m_d=m_s=0}\,. 
$$
Using the three-flavour to two-flavour matching relations~\cite{Gasser:1984gg}
\bea 
F &=& F_0\left (1-\frac{{\bar M}_K^2}{32\pi^2F_0^2}\,\ln\frac{{\bar M}_K^2}{\mu^2}+\frac{8{\bar M}_K^2}{F_0^2}\,L_4^r\right )\,, \no 
B &=& B_0\left [1-\frac{{\bar M}_{\eta}^2}{96\pi^2F_0^2}\,\ln\frac{{\bar M}_{\eta}^2}{\mu^2}+\frac{16{\bar M}_K^2}{F_0^2}\,
        \left (2L_6^r-L_4^r\right )\right ]\,, \nonumber
\eea
with
$$
{\bar M}_K^2\,=\,B_0m_s\,, \quad  {\bar M}_{\eta}^2\,=\,\frac{4}{3}\,B_0m_s\,, 
\quad B_0\,=\,-\frac{\langle{\bar u}u\rangle}{F_0^2}\biggm |_{m_u=m_d=m_s=0}\,,
$$
together with the three-flavour~\cite{Urech:1994hd} and two-flavour~\cite{Knecht:1997jw} calculations of $\Delta_{\pi}$ we get
$$
Z\,=\,Z_0\left (1-\frac{32{\bar M}_K^2}{F_0^2}\,L_4^r\right )+\frac{4{\bar M}_K^2}{F_0^2}\,K_8^r\,.
$$
Note that the following replacements are valid at the order we are working  
$$
Z\,=\,Z_0\rightarrow \frac{\Delta_{\pi}}{2e^2\Fpi}\,, \qquad B_0m_s\rightarrow \MKc -\frac{1}{2}\,\Mpic\,,
$$
with $\mKc =493.677\,\textrm{MeV}$. Using the values of the $K_i^r$'s derived in~\cite{Baur:1996ya} at $\mu =M_{\rho}=770\,\textrm{MeV}$ 
and assigning to each of them an uncertainty of $\pm 1/(16\pi^2)$ coming from na\"{\i}ve dimensional analysis, we obtain for the 
combinations (\ref{eqa:matching}) and (\ref{eqb:matching}) 
\be \label{eq:numK}
\frac{e^2\Mpin}{\Fpi}\,{\cal K}^{+-;+-}\,=\,8.63\pm 12.02\,, \quad \frac{e^2\Mpin}{\Fpi}\,{\cal K}^{++;++}\,=\,-15.13\pm 14.62\,.
\ee
This entails (we first show separately the contributions of each of 
the three terms in Eqs. (\ref{eqa:expression for 
the electromagnetic corrections}) and (\ref{eqb:expression for 
the electromagnetic corrections}))
\bea
\Delta a_0(+-;+-) 
  &=& {2.9\cdot 10^{-3}} 
  \,+\, {(1.9\pm 0.9)\cdot 10^{-4}} \,+\,
        {(2.5\pm 7.6)\cdot 10^{-4}} \no
&=& 
(3.2\pm 0.8)\cdot 10^{-3}\,, \no 
\Delta a_0(++;++) 
  &=& {2.9\cdot 10^{-3}} 
  \,+\, {(8.7\pm 1.1)\cdot 10^{-4}} \,+\,
        {(-12.0\pm 9.2)\cdot 10^{-4}} \no
 &=& (2.5\pm 0.9)\cdot 10^{-3}\,. 
\eea         
From the last result, we see that the size of electromagnetic corrections to 
the combination $2a_0^0+a_0^2$ represents $\sim 5\%$ 
of the NLO value. This amounts to one-half 
of the size of the next-to-next-to-leading order (NNLO) strong
interaction corrections~\cite{Colangelo:2000jc}. Also, the bulk of the 
corrections arises from the electromagnetic pion mass difference, whereas the 
uncertainties are dominated by the error bars on the determinations of
${\cal K}^{+-;+-}$ and ${\cal K}^{++;++}$. 



\paragraph{}

{\bf {5.}}~In the present work, the scattering amplitude for the process 
$\pi^+\pi^-\rightarrow\pi^+\pi^-$ 
was calculated at LO and NLO in the chiral counting and in the presence of 
electromagnetic 
corrections. We have also evaluated the influence of the latter on the 
corresponding combination $2a_0^0+a_0^2$ of $S$-wave 
$\pi\pi$ scattering lengths which is relevant for the $2S-2P$ strong 
energy level shift of pionium. The LO 
electromagnetic corrections come entirely from the difference between the 
charged and neutral pions masses. As for the NLO 
electromagnetic corrections, their size amounts to $\sim 10\%$ of the one for the lowest order isospin breaking effects. 
This corresponds to a $5\%$ correction to the strong $2S-2P$ level shift at NLO.


\paragraph{}

We thank H.~Sazdjian for useful discussions.




\begin{thebibliography}{99}

\bibitem{Gasser:2001iz}
J.~Gasser, A.~Rusetsky and J.~Schacher,
``HadAtom01'', Workshop on Hadronic Atoms, Bern, October 11 - 12 2001,
hep-ph/0112293.

\bibitem{Adeva:1994xz}
B.~Adeva {\it et al.},
CERN-SPSLC-95-1. 

\bibitem{Sigg:qe}
D.~Sigg, A.~Badertscher, P.~F.~Goudsmit, H.~J.~Leisi and G.~C.~Oades,
Nucl.\ Phys.\ A {\bf 609} (1996) 310; \\
H.~J.~Leisi,
PiN Newslett.\  {\bf 15} (1999) 258; \\
H.~C.~Schroder {\it et al.},
Phys.\ Lett.\ B {\bf 469} (1999) 25; Eur.\ Phys.\ J.\ C {\bf 21} (2001) 473.\\
D.~Gotta,
PiN Newslett.\  {\bf 15} (1999) 276.

\bibitem{Breunlich:af}
W.~H.~Breunlich {\it et al.}  [DEAR Collaboration],
PiN Newslett.\  {\bf 15} (1999) 266; \\
M.~Augsburger {\it et al.},
Nucl.\ Phys.\ A {\bf 663} (2000) 561.

\bibitem{Iwasaki:1997wf}
M.~Iwasaki {\it et al.},
Phys.\ Rev.\ Lett.\  {\bf 78} (1997) 3067; \\
M.~Iwasaki {\it et al.},
Nucl.\ Phys.\ A {\bf 639} (1998) 501.

\bibitem{Adeva:2000vb}
B.~Adeva {\it et al.},
CERN-SPSC-2000-032. 

\bibitem{Deser:1954vq}
S.~Deser, M.~L.~Goldberger, K.~Baumann and W.~Thirring,
Phys.\ Rev.\  {\bf 96} (1954) 774.

\bibitem{Bilenkii:zd}
S.~M.~Bilenkii, V.~H.~Nguyen, L.~L.~Nemenov and F.~G.~Tkebuchava,
Yad.\ Fiz.\  {\bf 10} (1969) 812; \\
J.~Uretsky and J.~Palfrey, 
Phys.\ Rev.\ {\bf 121} (1961) 1798.




\bibitem{Colangelo:2000jc}
G.~Colangelo, J.~Gasser and H.~Leutwyler,
Phys.\ Lett.\ B {\bf 488} (2000) 261.

\bibitem{Moor:ye}
U.~Moor, G.~Rasche and W.~S.~Woolcock,
Nucl.\ Phys.\ A {\bf 587} (1995) 747.

\bibitem{Gashi:1997ck}
A.~Gashi, G.~Rasche, G.~C.~Oades and W.~S.~Woolcock,
Nucl.\ Phys.\ A {\bf 628} (1998) 101.



\bibitem{Jallouli:1997ux}
H.~Jallouli and H.~Sazdjian,
Phys.\ Rev.\ D {\bf 58} (1998) 014011
[Erratum-ibid.\ D {\bf 58} (1998) 099901]; \\
H.~Sazdjian,
Phys.\ Lett.\ B {\bf 490} (2000) 203. 

\bibitem{Lyubovitskij:1996mb}
V.~E.~Lyubovitskij and A.~Rusetsky,
Phys.\ Lett.\ B {\bf 389} (1996) 181; \\
V.~E.~Lyubovitskij, E.~Z.~Lipartia and A.~G.~Rusetsky,
Pisma Zh.\ Eksp.\ Teor.\ Fiz.\  {\bf 66} (1997) 747
[JETP Lett.\  {\bf 66} (1997) 783]; \\
M.~A.~Ivanov, V.~E.~Lyubovitskij, E.~Z.~Lipartia and A.~G.~Rusetsky,
Phys.\ Rev.\ D {\bf 58} (1998) 094024.

\bibitem{Kong:1998xp}
X.~Kong and F.~Ravndal,
Phys.\ Rev.\ D {\bf 59} (1999) 014031; \\
X.~W.~Kong and F.~Ravndal,
Phys.\ Rev.\ D {\bf 61} (2000) 077506; \\
B.~R.~Holstein,
Phys.\ Rev.\ D {\bf 60} (1999) 114030; \\
D.~Eiras and J.~Soto,
Nucl.\ Phys.\ Proc.\ Suppl.\  {\bf 86} (2000) 267
[PiN Newslett.\  {\bf 15} (2000) 181]; \\
D.~Eiras and J.~Soto,
Phys.\ Rev.\ D {\bf 61} (2000) 114027.


\bibitem{Gasser:2001un}
J.~Gasser, V.~E.~Lyubovitskij, A.~Rusetsky and A.~Gall,
Phys.\ Rev.\ D {\bf 64} (2001) 016008.

\bibitem{Gasser:1999vf}
J.~Gasser, V.~E.~Lyubovitskij and A.~Rusetsky,
Phys.\ Lett.\ B {\bf 471} (1999) 244.

\bibitem{Gall:1999bn}
A.~Gall, J.~Gasser, V.~E.~Lyubovitskij and A.~Rusetsky,
Phys.\ Lett.\ B {\bf 462} (1999) 335.

\bibitem{Knecht:1997jw}
M.~Knecht and R.~Urech,
Nucl.\ Phys.\ B {\bf 519} (1998) 329.

\bibitem{Nemenov:vp}
L.~L.~Nemenov and V.~D.~Ovsyannikov,
Phys.\ Lett.\ B {\bf 514} (2001) 247.

\bibitem{Efimov:1985fe}
G.~V.~Efimov, M.~A.~Ivanov and V.~E.~Lyubovitskij,
Sov.\ J.\ Nucl.\ Phys.\  {\bf 44} (1986) 296
[Yad.\ Fiz.\  {\bf 44} (1986) 460].

\bibitem{Eiras:2000rh}
D.~Eiras and J.~Soto,
Phys.\ Lett.\ B {\bf 491} (2000) 101.

\bibitem{Lyubovitskij:2000kk}
V.~E.~Lyubovitskij and A.~Rusetsky,
Phys.\ Lett.\ B {\bf 494} (2000) 9.

\bibitem{Gasser:1983yg}
J.~Gasser and H.~Leutwyler,
Annals Phys.\  {\bf 158} (1984) 142.

\bibitem{Meissner:1997fa}
U.~G.~Meissner, G.~Muller and S.~Steininger,
Phys.\ Lett.\ B {\bf 406} (1997) 154
[Erratum-ibid.\ B {\bf 407} (1997) 454].


\bibitem{Kubis:1999db}
B.~Kubis and U.~G.~Meissner,
Nucl.\ Phys.\ A {\bf 671} (2000) 332
[Erratum-ibid.\ A {\bf 692} (2000) 647].

\bibitem{Weinberg:1966kf}
S.~Weinberg,
Phys.\ Rev.\ Lett.\  {\bf 17} (1966) 616. 

\bibitem{Nemenov:1984cq}
L.~L.~Nemenov,
Sov.\ J.\ Nucl.\ Phys.\  {\bf 41} (1985) 629
[Yad.\ Fiz.\  {\bf 41} (1985) 980].

\bibitem{Belkov:1986xn}
A.~A.~Belkov, V.~N.~Pervushin and F.~G.~Tkebuchava,
Yad.\ Fiz.\  {\bf 44} (1986) 466
[Sov.\ J.\ Nucl.\ Phys.\  {\bf 44} (1986) 300].

\bibitem{Groom:in}
D.~E.~Groom {\it et al.}  [Particle Data Group Collaboration],
Eur.\ Phys.\ J.\ C {\bf 15} (2000) 1.

\bibitem{Bijnens:1998fm}
J.~Bijnens, G.~Colangelo and P.~Talavera,
JHEP {\bf 9805} (1998) 014.

\bibitem{Baur:1996ya}
R.~Baur and R.~Urech,
Nucl.\ Phys.\ B {\bf 499} (1997) 319.

\bibitem{Moussallam:1997xx}
B.~Moussallam,
Nucl.\ Phys.\ B {\bf 504} (1997) 381.

\bibitem{Bijnens:1996kk}
J.~Bijnens and J.~Prades,
Nucl.\ Phys.\ B {\bf 490} (1997) 239.

\bibitem{Nehme:2002}
A.~Nehme, unpublished and PhD. Thesis (in preparation).

\bibitem{Gasser:1984gg}
J.~Gasser and H.~Leutwyler,
Nucl.\ Phys.\ B {\bf 250} (1985) 465.

\bibitem{Urech:1994hd}
R.~Urech,
Nucl.\ Phys.\ B {\bf 433} (1995) 234.




\end{thebibliography}
\end{document}